\documentclass{iau} 
\usepackage{graphicx,bm}
\newcommand{\chandra}{{\it Chandra}}
\newcommand{\xmm}{{\it XMM-Newton}}

\newcommand{\Rsolar}{ R$_{\odot}$}

\title[Eclipses in the Wolf-Rayet ULX CG\,X-1] 
{Monitoring of the eclipsing Wolf-Rayet ULX in the Circinus galaxy}

\author[Yanli Qiu \& Roberto Soria]   
{Yanli Qiu$^1$
 \and Roberto Soria$^2$}

\affiliation{$^1$National Astronomical Observatories, Chinese Academy of Sciences, 20A Datun Rd, Beijing 100101, China
\\ email: {\tt qiuyanli824@163.com} \\[\affilskip]
$^2$College of Astronomy and Space Sciences, University of the Chinese Academy of Sciences, Beijing 100049, China \\email: {\tt rsoria@nao.cas.cn}}

\pubyear{2019}
\volume{346}  
\jname{High Mass X-ray Binaries: illuminating the passage from
massive binaries to merging compact objects}
\editors{L.~Oskinova, ed.}
\begin{document}

\maketitle

\begin{abstract}
We studied the eclipsing ultraluminous X-ray source CG\,X-1 in the Circinus galaxy, re-examining two decades of {\it Chandra} and {\it XMM-Newton} observations. The short binary period (7.21 hr) and high luminosity ($L_{\rm X} \approx 10^{40}$ erg s$^{-1}$) suggest a Wolf-Rayet donor, close to filling its Roche lobe; this is the most luminous Wolf-Rayet X-ray binary known to-date, and a potential progenitor of a gravitational-wave merger. We phase-connect all observations, and show an intriguing dipping pattern in the X-ray lightcurve, variable from orbit to orbit. We interpret the dips as partial occultation of the X-ray emitting region by fast-moving clumps of Compton-thick gas. We suggest that the occulting clouds are fragments of the dense shell swept-up by a bow shock ahead of the compact object, as it orbits in the wind of the more massive donor.

\end{abstract}

\firstsection 
\section{Introduction}

The Circinus galaxy, located at a distance of 4.2 Mpc (\cite[Tully et al. 2009]{tully09}), contains a bright, point-like X-ray source known as CG\,X-1, seen at a projected distance of $\approx$300 pc from its starburst nucleus (Figure 1). The interpretation of this source has been the subject of debate for the past two decades (\cite[Bauer et al. 2001]{bauer01}; \cite[Weisskopf et al. 2004]{weisskopf04}; \cite[Esposito et al. 2015]{esposito15}). There are at least three features that make this object interesting and unusual. The first one is its high luminosity. If it is indeed located inside the Circinus galaxy (rather than being a foreground or background object), its average X-ray luminosity would be $\approx$10$^{40}$ erg s$^{-1}$; this would place CG\,X-1 near the top of the luminosity distribution of ultraluminous X-ray sources (ULXs) in the local universe: a factor of 10 times above the Eddington luminosity of Galactic stellar-mass black holes (BHs), or 100 times above the Eddington limit of a neutron star (NS). When CG\,X-1 was first discovered (\cite[Bauer et al. 2001]{bauer01}), observational and theoretical understanding of super-Eddington X-ray binaries was still in its infancy; however, today we know such sources exist and we can quantify their population properties. Based on the star formation rate of Circinus ($\approx$3--8 $M_{\odot}$ yr$^{-1}$: \cite[For et al. 2012]{for12}) and on the X-ray luminosity function of \cite[Mineo et al. (2012)]{mineo12}, we expect $\approx$0.2--0.6 X-ray binaries with a luminosity of 10$^{40}$ erg s$^{-1}$ or above, in that galaxy. Alternative interpretations, for example that of a foreground CV (\cite[Weisskopf et al. 2004]{weisskopf04}), can be rejected based on its high X-ray/optical flux ratio (\cite[Bauer et al. 2001]{bauer01}; Qiu et al. 2019 submitted), as well as the low probability of finding a foreground Galactic source projected onto the star-forming nucleus of Circinus.

The second interesting property of CG\,X-1 is the periodicity identified in its X-ray lightcurve. Most ULXs show stochastic variability by a factor of a few; in a few cases, periodic signals of a few days ({\it e.g.}, $\approx$2.5 d in M\,82 X-2: \cite[Bachetti et al. 2014]{bachetti14}; either $\approx$6 d or $\approx$13 d in M\,51 ULX-1: \cite[Urquhart \& Soria 2016a]{urquhart16a}; $\approx$8.2 d in M\,101 ULX-1: \cite[Liu et al. 2013]{liu13}) or even weeks ($\approx$64 d for NGC\,7793 P13: \cite[Motch et al. 2014]{motch14}) have been identified and interpreted as the binary periods. 
CG\,X-1 stands out with an unambiguous X-ray period of only 7.2 hr, determined from its eclipsing behaviour (\cite[Esposito et al. 2015]{esposito15}). This period is too short to be consistent with a supergiant donor; it suggests instead a Wolf-Rayet donor, whose radius is small enough to fit into such a compact binary system. In any case, CG\,X-1 is the ULX with the most precisely known binary period.

The third intriguing feature of CG\,X-1 is the nature of its X-ray eclipses. Although the period is stable (X-ray lightcurves observed twenty years ago can still easily be phase connected with recent observations) and the phase-averaged lightcurve is superficially consistent with the eclipse of the compact object behind the donor star, our study of the individual cycles tells a different story. Each cycle has a different pattern of eclipse duration and dipping morphology.

In this work, we will focus on the second and third property outlined above (short period and eclipse behaviour). We will discuss the general properties of the source in and out of eclipse, and the physical origin of the eclipses.  We will then briefly discuss the possible origin and future evolution of this system, and how its existence compares with the detection rate of gravitational wave events.

\begin{figure}
\begin{center}
 \includegraphics[width=6.25cm]{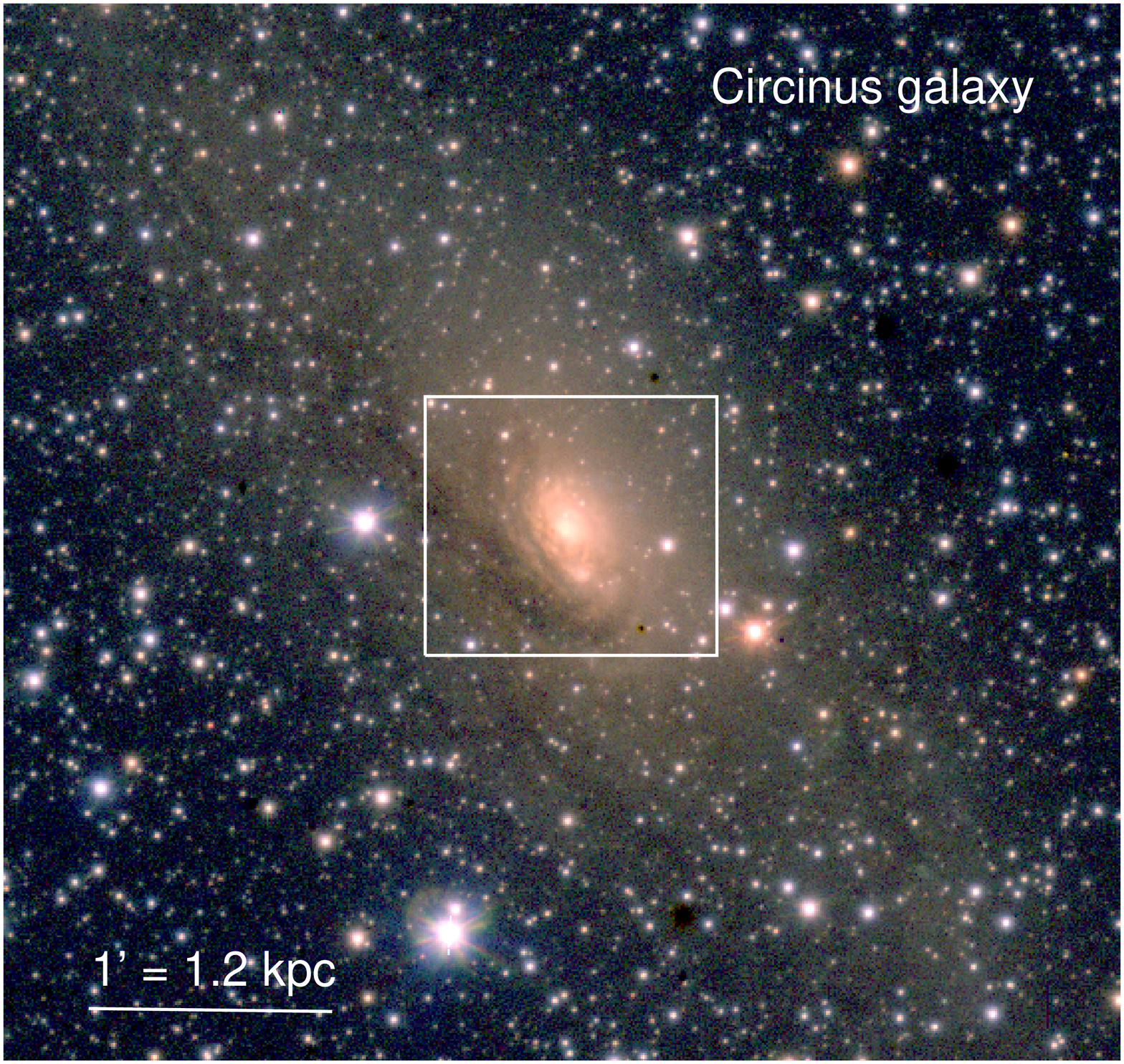} 
 \includegraphics[width=6.75cm]{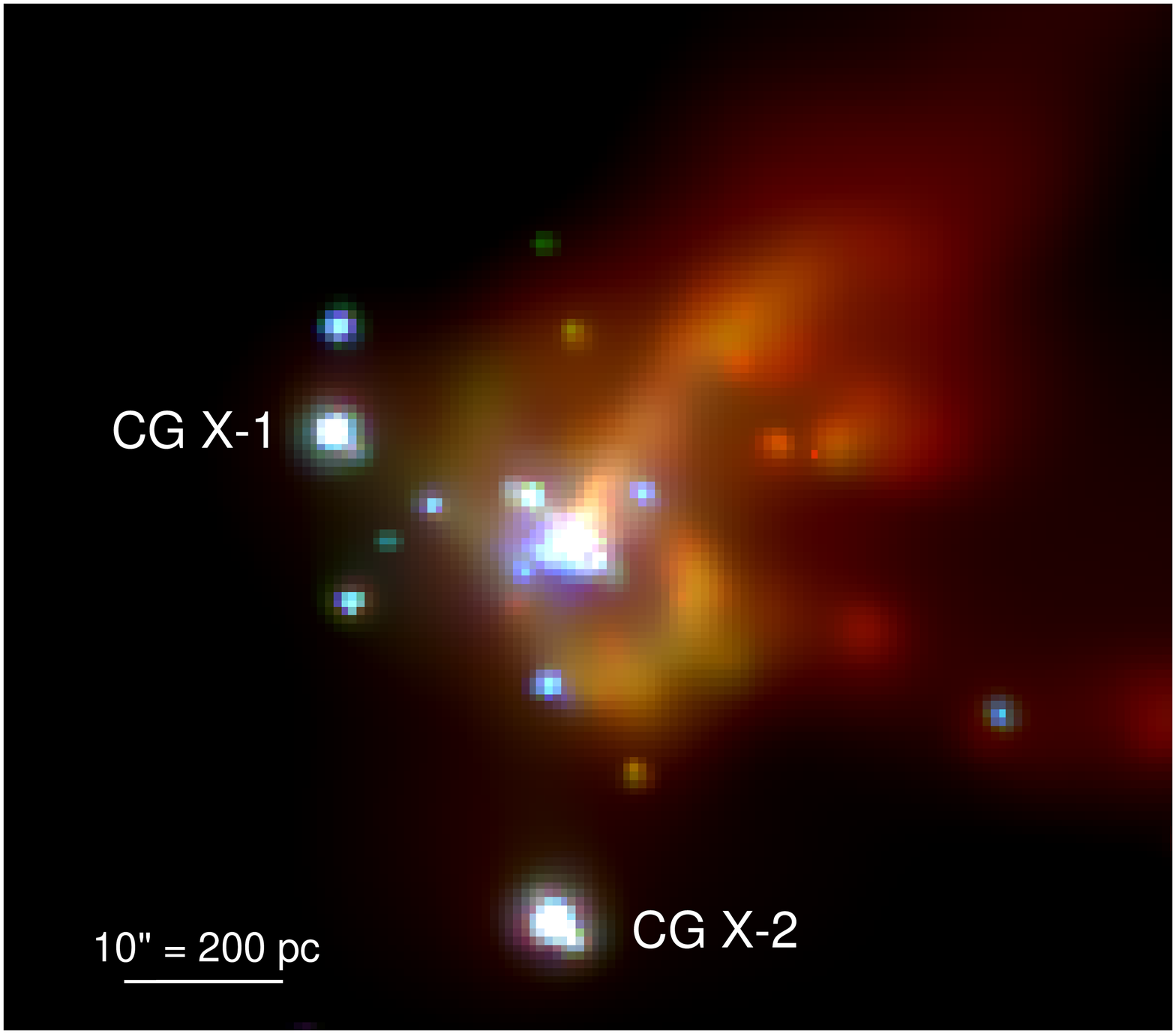} 
 \caption{Left panel: true-colour optical image of the Circinus galaxy in the {\it g,r,i} filters; data from the 2.6-m ESO-VLT Survey Telescope.  North is up and east to the left. The white
box is the area displayed in the {\it Chandra} image. Right panel: adaptively smoothed {\it Chandra}/ACIS image of the innermost region of the galaxy; red: 0.3--1.1 keV, green: 1.1--2.0 keV, blue: 2.0--7.0 keV. CG\,X-1 and  CG\,X-2 (also known as SN 1996cr: \cite[Bauer et al. 2008]{bauer08}) are the two brightest non-nuclear X-ray sources in the nuclear region of Circinus. A powerful outflow of hot thermal plasma from the nuclear star-burst is well visible to the west of the nucleus, but the soft diffuse emission is absorbed by thick dust to the east of the nucleus.}
   \label{fig1}
\end{center}
\end{figure}

\section{The most luminous Wolf-Rayet ULX}
There is an unresolved optical counterpart detected in the only (short) {\it Hubble Space Telescope} observation of the field, within the {\it Chandra} error circle for CG\,X-1. We estimate an apparent brightness $m_{\rm F606W} \approx V \approx 24.3 \pm 0.1$ mag in the Vega system; the distance modulus of Circinus is $\approx 28.1$ mag. Unfortunately, the optical extinction is very high, because Circinus is located behind the disk of the Milky Way; the line-of-sight extinction towards the Circinus galaxy is $A_V \approx 4$ mag (\cite[Schlafly \& Finkbeiner 2011]{schlafly11}), but additional local extinction is likely very high (see also \cite[Weisskopf et al. 2004]{weisskopf04}) and uncertain, given the location of the source near dust lanes. This, coupled with the lack of observations in other optical/IR bands, makes it impossible to determine the nature of this object (let alone its optical time variability). Thus, the main constraints to the nature of the system must come from the X-ray data.

\begin{table}
  \begin{center}
  \caption{Summary of candidate Wolf-Rayet X-ray binaries, in order of increasing period. }
  \label{tab1}
 {\scriptsize
  \begin{tabular}{|l|c|c|c|c|c|}\hline 
{\bf Name} & {\bf Galaxy} & {\bf Distance} & {\bf Peak $L_{0.3-10}^a$} & {\bf Period} & {\bf References} \\ 
   &  & {\bf (Mpc)} & {\bf (erg s$^{-1}$)} & {\bf hr} & \\ \hline
CXOU\,J121538.2$+$361921 & NGC\,4214 & 3.0 & $\approx$6$\times 10^{38}$ & 3.6 & 1\\[2pt]
Cygnus X-3 & Milky Way & 0.0074 & $\approx$ a few $\times 10^{38}$ & 4.8 & 2,3,4,5\\[2pt]
CXOU\,J123030.3$+$413853 & NGC\,4490 & 6.4 & $\approx$1$\times 10^{39}$ & 6.4 & 6\\ [2pt]
{\bf CG\,X-1} & {\bf Circinus} & {\bf 4.2}  &  $\bm{\approx}${\bf{3}}$\bm{\times 10^{40}}$ & {\bf 7.2} & {\bf 7,8}\\ [2pt]
CXOU\,J004732.0$-$251722 & NGC\,253 & 3.2 & $\approx$1$\times 10^{38}$ & 14.5 & 9\\ [2pt]
CXOU\,J005510.0$-$374212 (X-1) & NGC\,300 & 1.9 & $\approx$3$\times 10^{38}$ &   32.8 & 10,11,12,13 \\[2pt]
CXOU\,J002029.1$+$591651 (X-1) & IC\,10 & 0.7 & $\approx$7$\times 10^{37}$ &   34.8 & 14,11,15,16 \\
\hline
$^b$CXOU\,J140332.3$+$542103 (ULX-1) & M\,101 & 6.4 & $\approx$4$\times 10^{39}$ & 196.8 & 17,18,19\\
\hline
  \end{tabular}
  }
 \end{center}
\vspace{1mm}
 \scriptsize{
 {\it References:}\\
1: \cite[Ghosh et al. (2006)]{ghosh06};
2: \cite[Hjalmarsdotter et al. (2009)]{Hjalmarsdotter09}; 
3: \cite[Koljonen et al. (2010)]{koljonen10}; 
4: \cite[Zdziarski et al. (2012)]{zdziarski12}; 
5: \cite[McCollough et al. (2016)]{mccollough16};
6: \cite[Esposito et al. (2013)]{esposito13};
7: \cite[Esposito et al. (2015)]{esposito15};
8: Qiu et al., 2019 submitted.;
9: \cite[Maccarone et al. (2014)]{maccarone14};
10: \cite[Carpano et al. (2007)]{carpano07};
11: \cite[Barnard et al. (2008)]{barnard08};
12: \cite[Crowther et al. (2010)]{crowther10};
13: \cite[Binder et al. (2011)]{binder11};
14: \cite[Prestwich et al. (2007)]{prestwich07}
15: \cite[Silverman \& Filippenko (2008)]{silverman08}
16: \cite[Laycock et al. (2015)]{laycock15}
17: \cite[Kong et al. (2004)]{kong04};
18: \cite[Liu et al. (2013)]{liu13}; 
19: \cite[Urquhart \& Soria (2016b)]{urquhart16b}.\\   
{\it Notes:}\\
$^a$De-absorbed 0.3--10 keV luminosity in the bright phase of the orbital cycle; values taken from the references listed in this Table, but rescaled to the distance adopted here.\\
  $^b$M\,101 ULX-1 differs from the other seven sources because it is an ultraluminous supersoft source, it does not show eclipses, and its binary separation is too large to permit a BH-BH merger in a Hubble time. \\
}
\end{table}

From our study of all the archival {\it Chandra} and {\it XMM-Newton}, and two {\it ROSAT} observations between 1997 March and 2018 February (see Qiu et al., 2019 submitted, for a detailed log), we derive an average binary period $P = 25,970.1 \pm 0.4$ s $\approx 7.214$ hr. Using the period-density relation for binary systems (\cite[Eggleton 1983]{eggleton83}), we obtain an average density of $\rho \approx 1.2 \rho_{\odot} \approx 1.7$ g cm$^{-3}$ inside the Roche lobe of the donor star, for a mass ratio $M_2/M_1 = 2$, or $\rho \approx 0.74 \rho_{\odot} \approx 1.1$ g cm$^{-3}$, for $M_2/M_1 = 10$.  This range of values already rules out main-sequence OB stars (in fact, any main-sequence star more massive than $\approx$1 $M_{\odot}$), blue supergiants, red supergiants or red giants. The persistent nature of the X-ray source over at least 20 years, and its location in a highly star-forming region, strongly suggest a young system with a donor star more massive than the compact object. A Wolf-Rayet star is consistent with all those constraints. If CG\,X-1 contains a 20-$M_{\odot}$ Wolf-Rayet star and a 10-$M_{\odot}$ BH, the binary separation is $\approx$5.8 $R_{\odot}$ and the size of the Roche Lobe of the star is $\approx$2.6 $R_{\odot}$: this is large enough to contain a Wolf-Rayet but not any other type of massive star. Cygnus X-3 is the prototypical example of a high-luminosity X-ray binary with a very short binary period, fed by a Wolf-Rayet star. Very few such systems are known to-date (Table 1). Understanding their formation and evolution has been one of the most important recent developments in the field of X-ray binaries (\cite[van den Heuvel 2019]{vandenHeuvel2019}).

\begin{figure}
\center
\includegraphics[width=8.5cm]{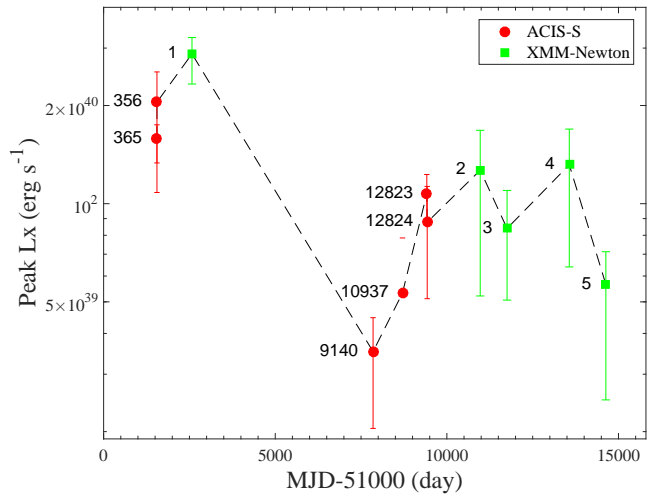}
\caption{0.3--8 keV lightcurve of CG\,X-1 between 2000 March 13 and 2018 February 7; we plot here the X-ray luminosity measured in the bright (non-occulted) phase of the observations (see Qiu et al. 2019, submitted, for more details).  Error bars are 90\% confidence levels. Red dots: \chandra/ACIS observations; green squares: \xmm/EPIC observations.  ObsIDs are labelled in the plot. }
\label{fig:long_lc}
\end{figure}

From our spectral modelling, we found (Qiu et al., 2019 submitted) that the de-absorbed X-ray luminosity of CG\,X-1 is $\approx$10$^{40}$ erg s$^{-1}$ during the out-of-eclipse parts of the orbital cycle, with a variability range over two decades spanning between $L_{\rm X} \approx 2.9 \times 10^{40}$ erg s$^{-1}$ ({\it XMM-Newton} observation of 2001 August 6) and $L_{\rm X} \approx 3.5 \times 10^{39}$ erg s$^{-1}$ ({\it Chandra} observation of 2008 October 26) (see Figure \ref{fig:long_lc} for the long-term lightcurve).
Such extreme luminosity implies a mass accretion rate onto the compact object of at least $\approx$10$^{-6} M_{\odot}$ yr$^{-1}$, for a radiative efficiency $\eta  \approx 0.15$. In fact, the radiative efficiency is likely to be lower, scaling as $\eta \sim 0.1 (1 + \ln \dot{m})/\dot{m}$ for super-Eddington accretion, where $\dot{m}$ is the accretion rate in Eddington units (\cite[Shakura \& Sunyaev 1973]{ss73}; \cite[Poutanen et al. 2007]{poutanen07}). Moreover, the eclipsing and dipping behaviour suggests a high viewing angle, so that we cannot invoke geometric beaming of the emission via a polar funnel. Thus, it appears that the system is really accreting at least several times $10^{-6} M_{\odot}$ yr$^{-1}$. Such high accretion rate suggests that the donor star is either filling the Roche lobe, or at least that its wind is gravitationally focused towards the compact object. Note that CG\,X-1 is the only system among candidate Wolf-Rayet X-ray binaries with a luminosity $\sim$10$^{40}$ erg s$^{-1}$; all others have X-ray luminosities $\lesssim$10$^{39}$ erg s$^{-1}$, consistent with wind accretion.

\begin{figure}
\hspace{-0.6cm}
\includegraphics[width=15.3cm]{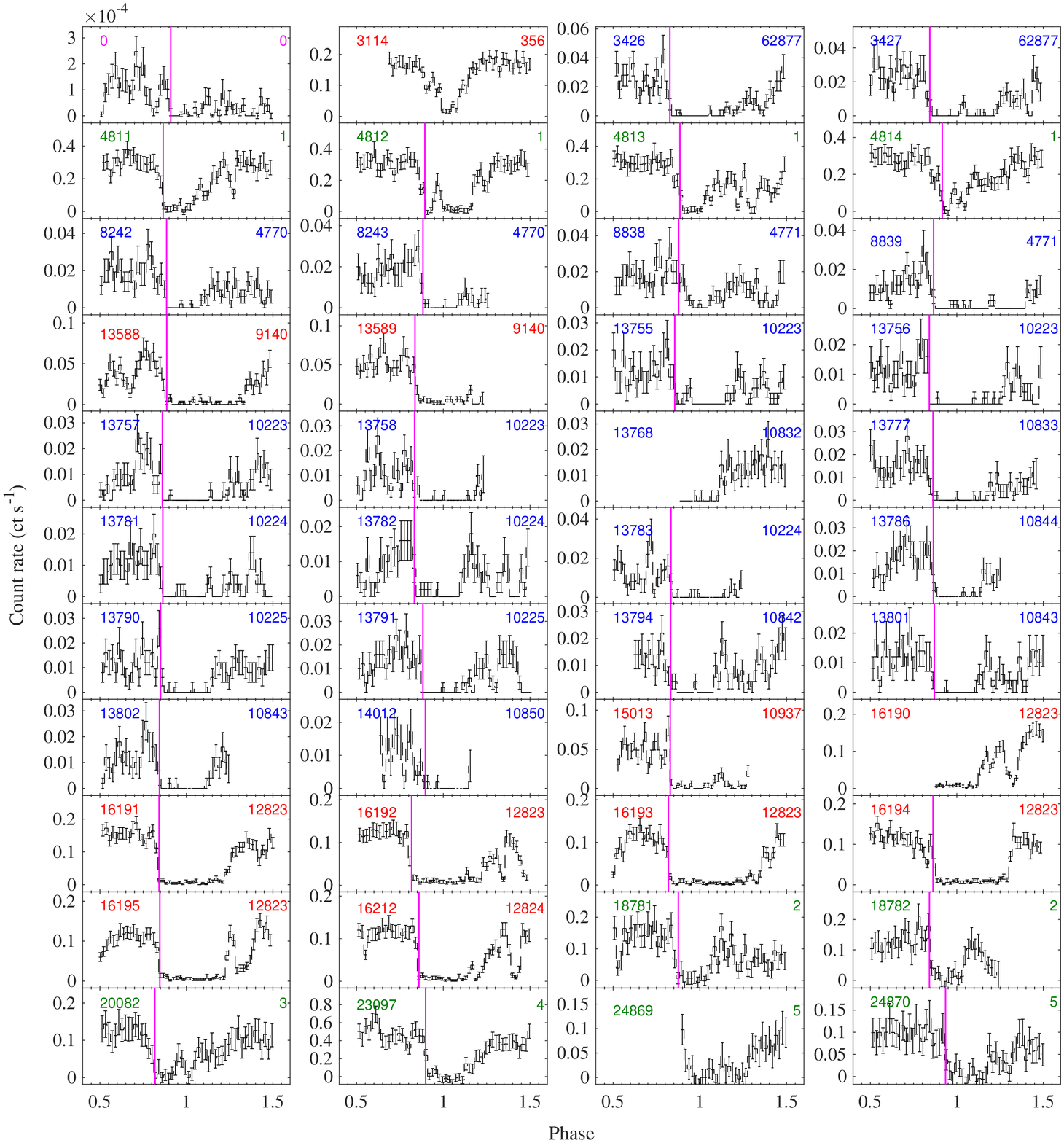} 
 \vspace{-0.7cm}
 \caption{ Twenty years of X-ray lightcurves for CG\,X-1, showing that each orbital cycle is different than the others, although some general underlying features remain the same. All lightcurves (apart from the one from {\it ROSAT}/HRI) are in the 0.3--8 keV band, and are binned to 500 s per bin.
  All lightcurves are folded on our best-fitting ephemeris; phase $\phi = 1$ occurs at MJD $ 50681.437036 + 0.300579 \times N$ (d). $N$. See Qiu et al.\ (2019, submitted) for details of how the ingress time and the ephemeris were calculated. In each panel, the number on the top left corresponds to the value of $N$ for the orbital cycle represented in that panel;  the number on the top right is a short form of the corresponding observation ID from which that lightcurves was extracted (see Qiu et al.\ 2019, submitted, for the full list of observations). The vertical magenta bar in each panel shows the estimated  mid-time of the eclipse ingress phase for that observation.
 The top left panel (labelled 0) is a {\it ROSAT}/HRI lightcurve from 1997; panels with red numbers correspond to {\it Chandra}/ACIS-S3 observations; those with blue numbers to {\it Chandra}/HETG observations; those with green numbers to {\it XMM-Newton}/EPIC observations. In the {\it XMM-Newton} panels, ObsID numbers 1, 2, 3, 4, 5 mean 0111240101, 0701981001, 0656580601, 0792382701, 0780950201, respectively.}
 %
 %
   \label{figalc}
\end{figure}

\section{Periodic eclipses and  dipping behaviour}

Folded X-ray lightcurves (Fig 6 in \cite[Esposito et al. 2015]{esposito15}; Qiu et al., 2019 submitted) show a sharp flux drop in ingress, followed by an eclipse phase lasting for $\approx$1/5 of the period, with faint residual emission (softer than out-of-eclipse), and then a slow return to the baseline flux. This structure is reminiscent of other high-mass X-ray binaries seen at high inclination, with a proper eclipse (apart from residual scattered photons) when the accreting object is behind the donor star, and varying absorbing column density at other phases, as the compact object moves through the wind of the donor. However, an inspection of the individual lightcurves of CG\,X-1 from each observed cycle tells a more complicated story (Figure \ref{figalc}). Any model of the system must explain the following two X-ray properties: \\
{\it a) the eclipse and dipping patterns and duration of the egress phase change from orbit to orbit.} Clearly, the size of the donor star and the binary separation cannot change; therefore, the eclipse and the dips must be (at least partly) caused by optically thick material ({\it e.g.}, clouds) in front of the X-ray source, moving on timescales shorter than the binary period.\\
{\it b) the transition from full eclipse to full flux level consists of a decreasing level of partial covering by an optically thick medium.} A sequence of X-ray spectra show (Qiu et al.\ 2019, submitted) that during the egress phase, the intrinsic spectral shape and cold absorption change only slightly, and the main difference is the value of the normalization constant. In other words, the flux recovery is not caused by a gradual decrease of absorbing column density. Instead, we propose that Compton-thick clouds ($N_{\rm H} > 1.5 \times 10^{24}$ cm$^{-2}$) occult a variable fraction (between 0 and 100\%) of the emitting region at different phases during each orbital cycle.

The occulting clouds cannot be uniformly or randomly distributed along all azimuthal angles, because the observed pattern of fast ingress, total eclipse, and dips during the slow egress is regularly repeated for two decades. In some low-mass X-ray binaries seen at high inclination, regular dipping behaviour is also observed, probably caused by the thick bulge where the accretion streams impacts the disk (\cite[White \& Swank 1982]{white82}; \cite[Frank et al. 1987]{frank87}). However, this scenario does not work for CG\,X-1, because the accretion stream always trails the compact object, and would produce dips just before or during eclipse ingress, contrary to the observed pattern. High-mass X-ray binaries sometimes also have an asymmetric eclipse profile due to a thick accretion stream ({\it e.g.},, Vela X-1: \cite[Doroshenko et al. 2013]{doroshenko13}), and in those cases, too, a slow ingress is followed by a fast egress. 

Taking those constraints into account, we suggest that the optically thick material is located between the Wolf-Rayet and the compact object, but mostly in front of the compact object. This configuration will lead to partial occultations of the X-ray emission after the compact object has passed behind the star and is moving towards us (egress), rather than before. 

We know that ULXs have strong radiatively driven winds ({\it e.g.}, \cite[Poutanen et al.\ 2007]{poutanen07}, \cite[Ohsuga \& Mineshige 2011]{ohsuga11}, \cite[Pinto et al.\ 2016]{Pinto2016}, \cite[Pinto et al.\ 2017]{Pinto2017}, \cite[Walton et al.\ 2016]{Walton2016}, \cite[Kosec et al.\ 2018]{Kosec2018}), probably comparable in kinetic energy
to Wolf-Rayet winds. We also estimate an orbital velocity $\approx$700 km s$^{-1}$ for a typical 10-$M_{\odot}$ stellar-mass BH orbiting a typical Wolf-Rayet star ($M_2 \approx 20$--30 $M_{\odot}$) on a 7.2-hr period. This implies that the compact object in GC\,X-1 is ploughing through the thick wind of the donor star at highly supersonic speed (Mach number $\mathcal{M} \approx 20$ for a wind temperature $\approx$10$^5$ K).

In general, in systems with a Wolf-Rayet star and an O star, a denser, optically-thick layer of shocked gas forms at the interface of the two winds; in CG\,X-1 we have the additional element that the compact object is moving at supersonic speed. We suggest that the compact object creates a bow shock along its direction of motion, and sweeps up a dense shell of shocked Wolf-Rayet wind; the density enhancement scales as $\mathcal{M}^2$, from standard bubble theory ({\it e.g.}, \cite[Weaver et al. 1977]{weaver77}). For typical ambient densities $n_{\rm e} \sim 10^{14}$ cm$^{-3}$ (\cite[Ro \& Matzner 2016]{ro16}) in the undisturbed wind, the density in the shocked shell can exceed $10^{16}$ cm$^{-3}$ and lead to rapid cooling. By analogy with Wolf-Rayet/O-star binaries (\cite[Usov 1991]{usov91}, \cite[Stevens et al.\ 1992]{stevens92}), a cold dust layer may form at the contact discontinuity between the shocked Wolf-Rayet wind and the shocked accretion-disk wind. For an order-of-magnitude estimate, we can assume a radius of the shell comparable to the Roche Lobe of the accreting object ($R \sim 10^{11}$ cm) and a thickness of the shell $\sim 10^8$ cm (\cite[Kenny \& Taylor 2005]{kenny05}): thus, the equivalent hydrogen column density in the swept-up shell can exceed $10^{24}$ cm$^{-2}$ and cause total occultation of the X-ray emission below 10 keV.

For high Mach numbers, hydrodynamic instabilities of the swept-up, cooling shell lead to continuous fragmentation and re-formation (\cite[Stevens et al.\ 1992]{stevens92}, \cite[Park \& Ricotti 2013]{park13}). We argue that fragments of the swept-up shell are the optically thick structures responsible for the irregular dipping in CG\,X-1. If this is the case, we expect to see occultations and dips in the X-ray lightcurve mostly when the compact object moves towards us (after egress from the eclipse), rather than before ingress.

\section{ Period derivative and binary evolution }

\begin{figure}
\center{
\includegraphics[width=8cm]{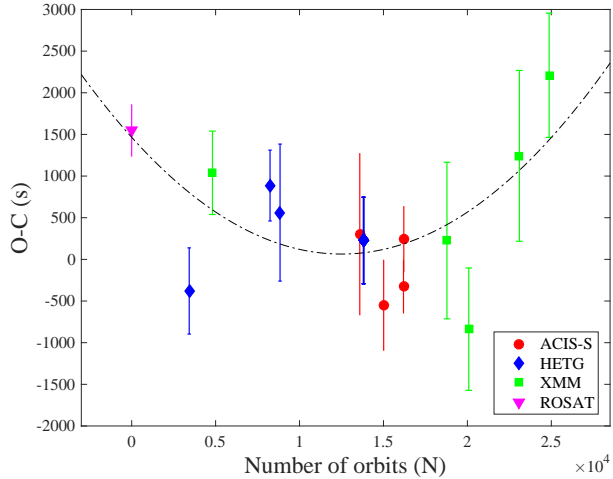}}
\caption{
The $O-C$ diagram for CG\,X-1, computed from the empirically determined ingress mid-times. The dashed black line is the best quadratic fit of the $O-C$ curve. The concave shape of the curve means that the period is increasing with time. Magenta datapoints are from the {\it ROSAT}/HRI observation; red ones from {\it Chandra}/ACIS observations; blue ones from {\it Chandra}/HETG observations; green ones from {\it XMM-Newton}/EPIC observations. }
\label{fig:oc}
\end{figure}

The eclipse ingress is a sharp feature generally recognizable in every orbital cycle (magenta lines in Fig. \ref{figalc}) and approximately periodic, but we do notice a significant dispersion in the phase of mid-ingress. This could be a random scatter caused by low-count statistics, or by some properties of the binary system ({\it e.g.}, variable optically thick outflows around the donor star). Alternatively, it could be a systematic drift caused by a period derivative $\dot{P} \neq 0$. In order to figure out whether the period is changing with time, we plotted an ``Observed minus Calculated'' ($O-C$) diagram. In an $O-C$ diagram, if the period remains constant with time, the datapoints follow a straight line; instead, if the period is changing linearly, the $O-C$ datapoints follow a quadratic function of time. 
In our case, the predicted time of mid-ingress $C$ is calculated from the folded lightcurve (stack of all \chandra\ and \xmm\ observations). The observed mid-time of ingress $O$ in each orbital cycle was determined with the method described in Qiu et al.\ (2019, submitted).
The resulting $O-C$ diagram clearly shows a concave curvature (Figure \ref{fig:oc}). This suggests that the orbital period and binary separation are systematically increasing. The period derivative calculated from the $O-C$ diagram is $\dot{P}/P = (8.4\pm5.0) \times 10^{-7}$ yr$^{-1}$ (significant at the 95\% confidence level). 

The increase of the orbital period is consistent with a huge mass loss rate from the binary system (non-conservative mass transfer), mainly caused by the Wolf-Rayet wind. For example, assuming for simplicity that all the mass lost by the system comes directly from the donor wind, $\dot{P}/P \approx -2 (\dot{M}_2/M_2)  \, q/(1+q)$ (\cite[Lommen et al.\ 2005]{lommen2005}). For representative values of $\dot{M}_2 \approx 10^{-5} M_{\odot}$ yr$^{-1}$, $M_2 \approx 20 M_{\odot}$, and $q \approx 3$ (typical of Wolf-Rayet stars and stellar-mass BHs), we do indeed expect $\dot{P}/P \approx 8 \times 10^{-7}$ yr$^{-1}$. Conversely, if most of the mass was transferred conservatively from the donor star to the Roche lobe of the compact object (via L1 overflow) and then ejected from the system via an accretion disk wind, the period change would be $\dot{P}/P \approx (\dot{M}_2/M_2)  \, (3q^2-2q-3)/(1+q)$ (\cite[Lommen et al.\ 2005]{lommen2005}), which is $<$0 (orbital shrinking) for $q \gtrsim  1.4$.

Short-period Wolf-Rayet X-ray binaries ($P_{\rm orb} \lesssim 1$d) are a rare and intriguing class of systems, both because they constrain an interesting phase of binary evolution, and because they may give rise to gravitational merger events in the near future. Such short periods indicate that they must have gone through a phase of substantial shrinking of the binary separation, from initial sizes of $\sim$100--1000 \Rsolar\ to $\lesssim$10 \Rsolar. 
The most likely mechanism for such extreme orbital shrinking is a common envelope phase after the formation of the first compact object ({\it {e.g.}}, \cite[]{Dominik2012}, \cite[Bogomazov et al. 2014]{Bogomazov2014}, \cite[Belczynski et al. 2016]{Belczynski2016}, \cite[van den Heuvel et al. 2017]{vandenHeuvel2017}, \cite[Bogomazov et al. 2018]{Bogomazov2018}, \cite[Giacobbo et al. 2018]{Giacobbo2018}). After the end of the common envelope phase, the massive donor star has lost all of its hydrogen envelope, and appears as a He core. If a Wolf-Rayet X-ray binary can survive and maintain a short binary separation after the second supernova explosion (collapse of the Wolf-Rayet star), its orbit will resume shrinking via gravitational wave emission, and the two compact objects will finally merge on a timescale shorter than the Hubble time ({\it {e.g.}}, \cite[Bulik et al. 2011]{Bulik2011}, \cite[Belczynski et al. 2013]{Belczynski2013}, \cite[Esposito et al. 2015]{esposito2015}, \cite[Belczynski et al. 2016]{Belczynski2016}). Therefore, an observational determination of the formation rate, lifetime, and volume density of compact Wolf-Rayet X-ray binaries in the local universe will provide crucial constraints to the rate of gravitational merger events (\cite[Abbott et al. 2016a]{Abbott2016a}, \cite[Abbott et al. 2016b]{Abbott2016b}, \cite[Abbott et al. 2017a]{Abbott2017a}, \cite[Abbott et al. 2017b]{Abbott2017b}, \cite[Abbott et al. 2017c]{Abbott2017c}). Because Wolf-Rayet X-ray binaries are rare, we need to look at a large volume of space to find a statistically representative sample. Our best chance to find more of these systems beyond a few Mpc is to search them in the ULX population (detectable at larger distances because of their super-Eddington luminosity). CG\,X-1 is the first example of a ULX that has been associated with a Wolf-Rayet donor; the majority of ULXs are instead consistent with supergiant donors, with periods of several days.

\section{Conclusions}

We have summarized some of the most interesting properties of the ULX CG\,X-1 in the Circinus galaxy. Its short orbital period (7.2 hr) makes it a strong candidate Wolf-Rayet X-ray binaries, a very rare system (less than 10 known to-date) with intriguing accretion physics. If the compact object is a stellar-mass BH and the Wolf-Rayet collapses into another BH (without disrupting the binary), the timescale for a gravitational merger is only $\approx$50 Myr (\cite[Esposito et al. 2015]{esposito15}). 
In addition, CG\,X-1 is one of the most luminous ULXs in the nearby universe, reaching peak luminosities in excess of $3 \times 10^{40}$ erg s$^{-1}$ at some epochs. By contrast, all other Wolf-Rayet X-ray binaries have luminosities $\lesssim$10$^{39}$ erg s$^{-1}$. Thirdly, the X-ray lightcurve shows a regular pattern of eclipses with fast ingress and slow egress (with a coherent phase over 20 years of observation), modified by an irregular pattern of deep dips, changing every orbital cycle. We have discussed a possible origin for the dips. We suggested that CG\,X-1 differs from sub-Eddington Wolf-Rayet systems such as Cyg X-3 because both the primary and the secondary launch a massive radiatively driven outflow. In fact, the gas environment in systems such as CG\,X-1 may be compared to binary Wolf-Rayet systems. 

In short, CG\,X-1 is an exceptional test case for studies of the progenitors of of gravitational wave events and their expected rate in the local universe; for studies of accretion and outflows in ULXs; and for studies of the hydrodynamics of colliding winds and shock-ionized bubbles. 

In a forthcoming paper (Qiu et al.\ 2019, submitted), we shall present a detailed X-ray timing and spectroscopic study of the system outside eclipse, in eclipse, and during egress. We will also discuss possible origins of the system (via a common envelope phase), various scenario for the occultations by optically thick clouds, and whether the accreting compact object is more likely to be a NS or a BH. 

\section*{Acknowledgements}
We thank Alexey Bogomazov, Jifeng Liu, Michela Mapelli, Manfred Pakull, Gavin Ramsay, Axel Schwope, Song Wang and Grzegorz Wicktorowicz for useful suggestions and discussions, which greatly improved our presentation at the IAU Symposium 346 in Vienna. We also thank Yu Bai, Rosanne Di Stefano, Dom Walton and Xiaojie Xu for their contributions as co-authors to the forthcoming ApJ paper. YQ acknowledges the Harvard-Smithsonian Center for Astrophysics, and RS thanks Curtin University and The University of Sydney, for hospitality during part of this research.

\end{document}